\documentclass[10pt,twocolumn,aps,showpacs,natbib,eqsecnum]{revtex4}

\usepackage{graphicx,color}
\usepackage{amssymb}

\def\ben{\begin{equation}}
\def\ene{\end{equation}}
\mathchardef\bigtilde="0365

\def\la{\langle}
\def\ra{\rangle}

\begin{document}

\title{%
Continuum Scaling from Large Mass Expansion on the Lattice:\\
Delta Expansion Applied to the Anharmonic Oscillator}

\author{Hideko Hashiguchi}\email{hashiguchi.hideko@it-chiba.ac.jp}
\affiliation{%
Division of Mathematics, Chiba Institute of Technology, Shibazono 2-1-1, Narashino, Chiba 275-0023, Japan}

\author{Keisuke Hoshino}\email{hoshino.keisuke@it-chiba.ac.jp}
\affiliation{%
Division of Mathematics, Chiba Institute of Technology, Shibazono 2-1-1, Narashino, Chiba 275-0023, Japan}
 
\author{Hirofumi Yamada}\email{yamada.hirofumi@it-chiba.ac.jp}
\affiliation{%
Division of Mathematics, Chiba Institute of Technology, Shibazono 2-1-1, Narashino, Chiba 275-0023, Japan}


\date{\today}

\begin{abstract}%
We dilate the scaling region of the lattice anharmonic oscillator at strong coupling by introducing the parameter $\delta$.  Performing expansion in $\delta$, the calculation of the mass gap in the continuum limit  via the series expansion effective at large lattice spacings is then studied.  We show that the dilation on the mass parameter $M$ recovers the scaling behavior of the hopping parameter $\beta$ and allows for precise approximation of the mass gap.   
\end{abstract}


\maketitle

\section{Introduction} 
Recently a new computational method is proposed 
to improve the scaling behavior of strong coupling expansion on the lattice \cite{yam} .  By the use of the method, scaling behavior of the ${\cal N}$ vector model at two dimension was reinvestigated by studying the relation between $\beta=\frac{1}{g}$ $(g:$ bare coupling constant) and the square of the dimensionless mass $M$ defined in the lattice momentum space.  

The new method starts with the series expansion generally available such as strong coupling expansion or hopping parameter expansion.  Specifically, $\beta$ is expanded in $1/M$ such that $\beta=\sum_{k=1}\frac{b_{k}}{M^k}$ and the function is dilated around $M=0$ by the change of variable, $M\to M(1-\delta)$ $(0\leq \delta\leq 1)$.   To access the continuum limit, the dilated $\beta$ at large $M$, $\beta(M(1-\delta))=\sum_{k=1}\frac{b_{k}}{M^k(1-\delta)^k}$, was expanded also in the dilation parameter $\delta$ and the value of $\delta$ is tuned to unity.  Since the series of $\beta$ is always truncated and expanded in $\delta$, there appears no divergence even $\delta$ is set to unity.   Then, in a wide region of new $M$, logarithmic scaling in accordance with the asymptotic freedom was found in truncated $1/M$ series of dilated $\beta$ in the large ${\cal N}$ limit.   At ${\cal N}=1$, rough scaling behavior of Ising model was also found.  

The proposed method has some similarities with the so called delta expansion \cite{early}, \cite{dun}.  The similarities become clear when the method proposed in \cite{yam} is applied to the continuum models with explicit mass term.  Though the point of view from the dilation is not mentioned in the existing literatures, we thus use the term "delta expansion" to refer the method.   However we like to point out following differences between the two methods:   In the conventional delta expansion, the  action $S$ of interest is generalized by introducing a parameter $\delta$ to $S_{\delta}=S_{0}+\delta(S-S_{0})$ where $S_{0}$ represents some solvable one.  Here $\delta$ is introduced as an interpolation parameter of $S_{0}$ and $S$.  Physical quantities are then expanded in $\delta$ and non-trivial results emerge after the substitution $\delta=1$.  However, in \cite{yam}, $\delta$ is introduced as the parameter to dilate the scaling region of the system described by $S$ itself.   Although the conventional delta expansion requires good choice of  $S_{0}$ and critical use of the principle of minimum sensitivity \cite{pms}, in the new delta expansion, $S_{0}$ is obsolete and the later plays less important role.   
  
  The purpose of the present paper is to apply the new version of delta expansion to the anharmonic oscillator at strong coupling.  A detailed study is presented on the subject of calculating the mass gap in the continuum limit, the gap energy between the ground state and the first excited state, via the series expansion effective at large lattice spacings.    At strong coupling, the mass gap is generated non-perturbatively since it depends on the quartic coupling constant $\lambda$ as $const\times \lambda^{1/3}$.  Thus the present investigation serves us a good testing ground of the new method.  
  
\section{Anharmonic oscillator and hopping expansion}
To focus on essential aspects, we confine ourselves with the pure anharmonic case where the harmonic mass term is absent from the action.  The action $S$ on the lattice with lattice spacing $a$ is given by
$$
S=\sum_{n=-\bar L}^{\bar L}a\bigg[\frac{1}{2}\Big(\frac{\phi_{n+1}-\phi_{n}}{a}\Big)^2+\frac{\lambda}{4}\phi_{n}^4\bigg]
$$
where $n$ ($n=0,\pm 1,\pm 2,\cdots, \pm \bar L;\, L=2\bar L+1$) denotes a lattice site and the real field $\phi$ on the site $n$ is written as $\phi_{n}$.  

The action can be simplified by rescaling field variables.  Let the rescaled field $\varphi$ be defined by
$$
\varphi_{n}=\Big(\frac{a\lambda}{4}\Big)^{-1/4}\phi_{n}.
$$
Then, the action takes the following form,
\begin{equation}
S=\beta\sum(\varphi_{n}^2-\varphi_{n+1}\varphi_{n})+\sum \varphi_{n}^4,
\end{equation} 
where \cite{kappa}
\begin{equation}
\beta=\Big(\frac{4}{\lambda a^3}\Big)^{1/2}.
\label{beta}
\end{equation}
Note that the parameter $\beta$ is small when $a$ or $\lambda$ is large.  

For the purpose of the present work, we need expansion of correlation length, or mass in momentum space in powers of $\beta$ and we use a technique of hopping expansion \cite{hop}.    As the first step we divide the action into the "potential" $\sum V$ and the hopping term $\beta\sum\varphi_{n+1}\varphi_{n}$ as
$$
S=\sum_{n} V(\varphi_{n})-\beta\sum_{n} \varphi_{n+1}\varphi_{n},\quad V(\varphi)=\beta \varphi^2+\varphi^4.
$$
An average of $\Omega$ is then calculated as follows:
\begin{eqnarray}
\la \Omega\ra&=&\frac{1}{Z}\int\prod_{i} \left[d\varphi_{i} 
e^{-V(\varphi_{i})}\right] \prod_{n}\exp\Big[\beta\varphi_{n+1}\varphi_{n}\Big]\,\Omega \nonumber\\
&=&\frac{1}{Z}\int\prod_{i} \left[d\varphi_{i} 
e^{-V(\varphi_{i})}\right] \times\bigg(1+\sum_{n}\beta\varphi_{n+1}\varphi_{n}\nonumber\\
& &+\sum_{m,n(m\neq n)}\beta^2(\varphi_{m+1}\varphi_{m})(\varphi_{n+1}\varphi_{n})\nonumber\\
& &+\sum_{n}\frac{\beta^2}{2!}(\varphi_{n+n}\varphi_{n})^2+\cdots\bigg)\Omega.
\end{eqnarray} 
Here $Z$ denotes the partition function given by
$$
Z=\int\prod_{i} d\varphi_{i} 
e^{-V(\varphi_{i})}\times\prod_{n}\exp\Big[\beta\varphi_{n+1}\varphi_{n}\Big],
$$
and $Z$ may be expanded in powers of the hopping term.

\subsection{Partition function to $\beta^{8}$}
Computation of expectation values includes the partition function as the divisor and we first compute $Z$ to $8$th order of $\beta$.  The hopping expansion of $Z$ to the first few orders reads
\begin{eqnarray}
Z&=&\int\prod_{i} \left[d\varphi_{i} 
e^{-V(\varphi_{i})}\right] \times\bigg(1+\sum_{n}\beta\varphi_{n+1}\varphi_{n}\nonumber\\
& &+\sum_{m,n(m\neq n)}\beta^2(\varphi_{m+1}\varphi_{m})(\varphi_{n+1}\varphi_{n})\nonumber\\
& &+\sum_{n}\frac{\beta^2}{2!}(\varphi_{n+n}\varphi_{n})^2+\cdots\bigg).\nonumber
\end{eqnarray} 
As shown in Figures 1 and 2, it is convenient to use a graphical representation of each contributions.
\begin{figure}[h]
\begin{center}
\includegraphics[scale=0.50]{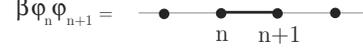}
\end{center}
\caption{Basic building block in hopping expansion.}
\end{figure}   
\begin{figure}[h]
\begin{center}
\includegraphics[scale=0.50]{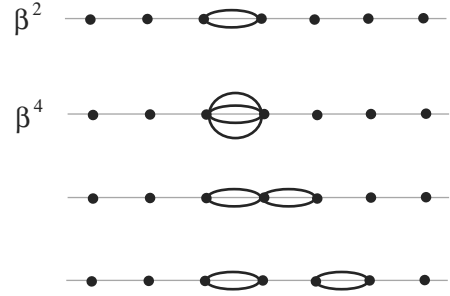}
\end{center}
\caption{The graphs contributing to the partition function to $4^{\rm th}$ order in the hopping term.}
\end{figure}   
To $\beta^{8}$, we obtain
\begin{eqnarray}
Z&=&h_{0}^{L}\bigg[1+L\beta^2\frac{\gamma_{1}^2}{2}+L\beta^4\Big(\frac{\gamma_{2}^2}{24}+\frac{\gamma_{1}^2\gamma_{2}}{4}+\frac{L-3}{12}\gamma_{1}^4\Big)\nonumber\\
& &+L\beta^6\Big((L-4)(L-5)\frac{\gamma_{1}^6}{48}-(L-4)\frac{\gamma_{1}^4\gamma_{2}}{8}\nonumber\\
& &+(L+3)\frac{\gamma_{1}^2\gamma_{2}^2}{48}+\frac{\gamma_{1}\gamma_{2}\gamma_{3}}{24}+\frac{\gamma_{3}^2}{720}\Big)\nonumber\\
& &+L\beta^8\Big(\frac{1}{384}(L-5)(L-6)(L-7)\gamma_{1}^8\nonumber\\
& &+\frac{1}{32}(L-5)(L-6)\gamma_{1}^6\gamma_{2}\nonumber\\
& &+\frac{1}{192}(L-5)(L+14)\gamma_{1}^4\gamma_{2}^2+\frac{1}{96}(L+2)\gamma_{1}^2\gamma_{2}^3\nonumber\\
& &+\frac{1}{1152}(L-3)\gamma_{2}^4+\frac{1}{48}(L-4)\gamma_{1}^3\gamma_{2}\gamma_{3}+\frac{1}{48}\gamma_{1}\gamma_{2}^2\gamma_{3}\nonumber\\
& &+\frac{1}{1440}(L+12)\gamma_{1}^2\gamma_{3}^2+\frac{1}{576}\gamma_{2}^2\gamma_{4}+\frac{1}{720}\gamma_{1}\gamma_{3}\gamma_{4}\nonumber\\
& &+\frac{1}{40320}\gamma_{4}^2\Big)+O(\beta^{10})\bigg],
\end{eqnarray}
where
\begin{eqnarray*}
h_{0}&=&\int_{-\infty}^{\infty}d\varphi e^{-V(\varphi)},\\
\gamma _{j}&=&\frac{1}{h_{0}}\int_{-\infty}^{\infty}d\varphi e^{-V(\varphi)} \varphi^{2j}=\frac{1}{h_{0}}\Big(-\frac{\partial }{\partial\beta}\Big)^{j}h_{0}.
\end{eqnarray*}
To complete the expansion, we must expand $\gamma_{j}$ in $\beta$.  The result of the expansion is written in Appendix A and, using the result (\ref{gamma}), we have the full expansion of $Z$ in powers  of $\beta$.  Exponentiating the contributions, we thus obtain
\begin{eqnarray} 
Z
&=&h_{0}^{L}\exp \bigg[ L\bigg\{\frac{\rho ^2 \beta ^2}{2}+\left(\rho ^3-\frac{\rho }{4}\right) \beta
   ^3\nonumber\\
   & &+\left(\frac{9 \rho ^4}{8}-\frac{3 \rho
   ^2}{16}+\frac{13}{384}\right) \beta ^4+\left(\frac{\rho ^5}{2}+\frac{\rho
   ^3}{8}-\frac{\rho }{16}\right) \beta ^5\nonumber\\
   & &+\left(-\frac{5 \rho ^6}{6}+\frac{9 \rho ^4}{16}-\frac{31 \rho
   ^2}{320}+\frac{3}{256}\right) \beta ^6\nonumber\\
   & &+\left(-2
   \rho ^7+\frac{5 \rho ^5}{8}-\frac{\rho ^3}{120}-\frac{5 \rho }{384}\right) \beta
   ^7\nonumber\\
   & &+\left(-\frac{91 \rho ^8}{64}-\frac{21 \rho ^6}{64}+\frac{2141 \rho
   ^4}{7680}-\frac{233 \rho ^2}{5120}+\frac{395}{114688}\right) \beta ^8\nonumber\\
   & &+O(\beta^9)\bigg\}\bigg],
\end{eqnarray} 
where
$$
\rho:=\frac{\Gamma(3/4)}{\Gamma(1/4)}=0.337989\cdots.
$$
The log of $Z$ is exactly proportional to $L$, the total number of cites.

\subsection{Mass variables from the two point function at large separation}
To compute the mass gap in the continuum limit, we must address to the relation between $\beta$ and a variable relevant to the mass.  One of quantities of our concern is therefore the inverse of the correlation length $\xi$ and it is extracted from the two point correlation function, $\la\varphi_{0}\varphi_{n}\ra$ at $n\gg 1$.  We calculate the correlation function by the use of the hopping expansion and, collecting the result to $\beta^{n+8}$ in $n$, we obtain
\begin{equation}
\la\varphi_{0}\varphi_{n}\ra=(\beta\gamma_{1})^n(c_{0}+c_{1}n+c_{2}n^2+c_{3}n^3+\cdots)
\end{equation}
where the coefficient $c_{k}\,(k=0,1,2,\cdots)$ is given in Appendix B.  By the exponentiation, we find that all terms involving $n$ in the logarithm of $\la\varphi_{0}\varphi_{n}\ra$ are linear in $n$.   Thus the exponential decay at large $n$, $\la\varphi_{0}\varphi_{n}\ra\sim const \exp(-n/\xi)$, is explicitly confirmed.   Then we can write  $\la\varphi_{0}\varphi_{n}\ra\sim c_{0}\exp[(\log\beta\gamma_{1}+\frac{c_{1}}{c_{0}})n]$ at large $n$ and $\xi^{-1}=-\log\beta\gamma_{1}-\frac{c_{1}}{c_{0}}$.  
By expanding $c_{1}/c_{0}$ in $\beta$ and $\gamma_{j}$, we have the inverse of correlation length,
\begin{eqnarray}
\xi^{-1}&=&-\log (\beta\gamma_{1})+\beta^2\Big(\frac{\gamma_{1}^2}{2}-\frac{\gamma_{2}^2}{6\gamma_{1}^2}\Big)+\beta^4\Big(-\frac{3\gamma_{1}^4}{8}\nonumber\\
& &+\frac{\gamma_{1}^2\gamma_{2}}{4}+\frac{\gamma_{2}^2}{24}+\frac{\gamma_{2}^4}{24\gamma_{1}^4}-\frac{\gamma_{2}^2\gamma_{3}}{36\gamma_{1}^3}-\frac{\gamma_{3}^2}{120\gamma_{1}^2}\Big)\nonumber\\
& &+\beta^6\Big(\frac{5\gamma_{1}^6}{12}-\frac{\gamma_{1}^4\gamma_{2}}{2}+\frac{\gamma_{1}^2\gamma_{2}^2}{16}-\frac{5\gamma_{2}^6}{324\gamma_{1}^6}+\frac{\gamma_{1}\gamma_{2}\gamma_{3}}{24}\nonumber\\
& &+\frac{\gamma_{2}^4\gamma_{3}}{54\gamma_{1}^5}+\frac{\gamma_{3}^2}{720}-\frac{\gamma_{2}^2\gamma_{3}^2}{2160\gamma_{1}^4}-\frac{\gamma_{2}\gamma_{3}\gamma_{4}}{360\gamma_{}^3}-\frac{\gamma_{4}^2}{5040\gamma_{1}^2}\Big)\nonumber\\
& &+\beta^8 \bigg(
-\frac{35\gamma_1^8}{64}+\frac{15\gamma_1^6 \gamma_2}{16}
-\frac{35\gamma_1^4 \gamma_2^2}{96}+\frac{\gamma_1^2 \gamma_2^3}{48}\nonumber\\
& &
-\frac{\gamma_2^4}{384}
+\frac{35\gamma_2^8}{5184\gamma_1^8}-\frac{\gamma_1^3\gamma_2\gamma_3}{12}
+\frac{\gamma_1\gamma_2^2\gamma_3}{48}-\frac{5\gamma_2^6\gamma_3}{432\gamma_1^7}\nonumber\\
& &
+\frac{\gamma_1^2\gamma_3^2}{120}+\frac{\gamma_2^4\gamma_3^2}{288\gamma_1^6}
+\frac{\gamma_2^2\gamma_3^3}{6480\gamma_1^5}+\frac{\gamma_3^4}{9600\gamma_1^4}
+\frac{\gamma_2^2\gamma_4}{576}\nonumber\\
& &+\frac{\gamma_1\gamma_3\gamma_4}{720}
+\frac{\gamma_2^3\gamma_3\gamma_4}{540\gamma_1^5}
-\frac{\gamma_2\gamma_3^2\gamma_4}{2160\gamma_1^4}+\frac{\gamma_4^2}{40320}
-\frac{\gamma_2^2\gamma_4^2}{7560\gamma_1^4}\nonumber\\
& &
-\frac{\gamma_3^2\gamma_5}{14400\gamma_1^3}
-\frac{\gamma_2\gamma_4\gamma_5}{15120\gamma_1^3}
-\frac{\gamma_5^2}{362880\gamma_1^2}
\bigg)+\cdots.
\label{corr}
\end{eqnarray}
One can obtain full expansion of $\xi^{-1}$ by expanding $\gamma_{j}$ in $\beta$ and the result reads
\begin{eqnarray}
\xi^{-1}&=&-\log(\rho\beta)+\Big(\frac{1}{4\rho}-\rho\Big)\beta+\Big(\frac{1}{48\rho^2}-\frac{1}{4}\Big)\beta^2\nonumber\\
& &+\bigg(\frac{1}{48\rho}-\frac{\rho}{4}+\frac{2\rho^3}{3}\bigg)\beta^3\nonumber\\
& &+\bigg(-\frac{5}{6144\rho^4}+\frac{11}{768\rho^2}-\frac{7}{160}-\frac{3\rho^2}{16}+\frac{7\rho^4}{8}\bigg)\beta^4\nonumber\\
& &+\bigg(-\frac{3}{10240\rho^5}+\frac{23}{4608\rho^3}-\frac{37}{1920\rho}-\frac{29\rho}{480}+\frac{\rho^3}{8}\nonumber\\
& &+\frac{3\rho^5}{10}\bigg)\beta^5+\bigg(-\frac{43}{663552\rho^6}+\frac{41}{36864\rho^4}-\frac{33}{7168\rho^2}\nonumber\\
& &+\frac{17}{1920}-\frac{91\rho^2}{960}+\frac{9\rho^4}{16}-\rho^6\bigg)\beta^6+
\bigg(
-\frac{11}{1548288\rho^{7}}\nonumber\\
& &+\frac{19}{221184\rho^5}
+\frac{83}{645120\rho^{3}}-\frac{31}{10752\rho}
-\frac{5\rho}{384}-\frac{\rho^3}{160}\nonumber\\
& &
+\frac{5\rho^5}{8}-\frac{15\rho^7}{7}\bigg)\beta^7+
\bigg(
\frac{515}{339738624\rho^8}
-\frac{137}{2359296\rho^6}\nonumber\\
& &
+\frac{5771}{8257536\rho^4}
-\frac{15259}{5160960\rho^2}+\frac{211}{35840}-\frac{233\rho^2}{5120}\nonumber\\
& &
+\frac{719\rho^4}{2560}
-\frac{21\rho^6}{64}-\frac{99\rho^8}{64}
\bigg)\beta^8+O(\beta^9) \nonumber\\
&=&-\log(\rho\beta)+\sum_{k=1}^{\infty}z_{k}\beta^k.
\label{xibetascale}
\end{eqnarray}

On the lattice, the inverse of the correlation length and the mass in the momentum space are rather different objects.  Their relation can be known by using Fourier representation.  
The Fourier representation of $const\times\exp(-n/\xi)$ is given by $const\times \int_{-\pi}^{\pi}\frac{d\theta}{2\pi}\frac{e^{in\theta}}{M+2(1-\cos\theta)}$.  By computing the integral, we find that
\begin{equation}
M=2\cosh(\xi^{-1})-2.
\label{relation}
\end{equation}
Note that at large $\xi$,
\begin{equation}
M\sim \xi^{-2}
\label{mxi}
\end{equation}
as it should be.  On the other hand, at small $\xi$, $M\sim \exp(\xi^{-1})$ and their mutual relation is quite different from (\ref{mxi}).  This might cause a non-negligible difference in the results obtained by the delta expansion as in the case of the large ${\cal N}$ anharmonic oscillator \cite{cit}.  The actual results in the present case shall be discussed in the next section.  

To make ready for the next section,  we express $1/M$ as a power series of $\beta$ and then invert the series.   From (\ref{xibetascale}) and (\ref{relation}), we find
\begin{eqnarray}
\frac{1}{M}&=&\rho \beta+\Big(-\frac{1}{4}+3\rho^2\Big)\beta^2+\Big(\frac{1}{96\rho}-\rho+\frac{15\rho^3}{2}\Big)\beta^3\nonumber\\
& &
+\Big(\frac{1}{384\rho^2}+\frac{3}{32}-\frac{23\rho^2}{8}+\frac{33\rho^4}{2}\Big)\beta^4\nonumber\\
& &+\Big(\frac{5}{9216\rho^3}-\frac{1}{256\rho}+\frac{109\rho}{320}-\frac{109\rho^3}{16}+33\rho^5\Big)\beta^5\nonumber\\
& &+\Big(\frac{1}{12288\rho^4}-\frac{1}{1536\rho^2}-\frac{19}{1920}+\frac{143\rho^2}{160}-\frac{227\rho^4}{16}\nonumber\\
& &+\frac{123\rho^6}{2}\Big)\beta^6+\Big(\frac{1}{442368\rho^5}+\frac{1}{24576\rho^3}-\frac{53}{215040\rho}\nonumber\\
& &-\frac{133\rho}{3840}+\frac{627\rho^3}{320}-\frac{865\rho^5}{32}+\frac{873\rho^7}{8}\Big)\beta^7\nonumber\\
& &+
\bigg(
-\frac{7}{1769472\rho^6}
+\frac{23}{294912\rho^4}
-\frac{95}{258048\rho^2}\nonumber\\
& &+\frac{1}{43008}
-\frac{337\rho^2}{3840}+\frac{2453\rho^4}{640}
-\frac{777\rho^6}{16}+\frac{1503\rho^8}{8}
\bigg)\beta^8\nonumber \\
& &+
\bigg(-\frac{149}{84934656\rho^7}
+\frac{13}{393216\rho^5}
-\frac{5}{28672\rho^3}\nonumber\\
& &
+\frac{1243}{5160960\rho}
+\frac{1567\rho}{921600}
-\frac{479\rho^3}{2560}
+\frac{1111\rho^5}{160}\nonumber\\
& &
-\frac{5385\rho^7}{64}
+\frac{2547\rho^9}{8}
\bigg)\beta^9 +O(\beta^{10})\nonumber\\
&=&\sum_{k=1}^{\infty}m_{k}\beta^k.
\label{smallb}
\end{eqnarray}
Inverting the above expansion,  we arrive at
\begin{eqnarray}
\beta&=&\frac{1}{\rho M}+\Big(\frac{1}{4\rho^3}-\frac{3}{\rho}\Big)\frac{1}{M^2}+\Big(\frac{11}{96\rho^5}-\frac{2}{\rho^3}+\frac{21}{2\rho}\Big)\frac{1}{M^3}\nonumber\\
& &+\Big(\frac{1}{16\rho^7}-\frac{3}{2\rho^5}+\frac{49}{4\rho^3}-\frac{39}{\rho}\Big)\frac{1}{M^4}\nonumber\\
& &+\Big(\frac{85}{2304\rho^9}-\frac{291}{256\rho^7}+\frac{4181}{320\rho^5}-\frac{1079}{16\rho^3}+\frac{597}{4\rho}\Big)\frac{1}{M^5}\nonumber\\
& &+\Big(\frac{93}{4096\rho^{11}}-\frac{879}{1024\rho^9}+\frac{24617}{1920\rho^7}-\frac{15151}{160\rho^5}+\frac{5603}{16\rho^3}\nonumber\\
& &-\frac{2325}{4\rho}\Big)\frac{1}{M^6}+\Big(\frac{6325}{442368\rho^{13}}-\frac{47411}{73728\rho^{11}}+\frac{170223}{14336\rho^9}\nonumber\\
& &-\frac{88343}{768\rho^7}+\frac{197331}{320\rho^5}-\frac{56063}{32\rho^3}+\frac{18315}{8\rho}\Big)\frac{1}{M^7}\nonumber\\
& &+\bigg(\frac{2015}{221184\rho^{15}}-\frac{17575}{36864\rho^{13}}+\frac{13581439}{1290240\rho^{11}}\nonumber\\
& &-\frac{2727527}{21504\rho^9}+\frac{431429}{480\rho^7}-\frac{18709}{5\rho^5}+\frac{136641}{16\rho^3}\nonumber\\
& &-\frac{36351}{4\rho}\bigg)\frac{1}{M^8}+
\bigg(
 \frac{123743}{21233664\rho^{17}}-\frac{2471275}{7077888\rho^{15}}\nonumber\\
 & &
+\frac{6196391}{688128\rho^{13}}-\frac{84066931}{645120\rho^{11}}+\frac{7448421371}{6451200\rho^9}\nonumber\\
& &
-\frac{98237681}{15360\rho^7}+\frac{13816421}{640\rho^5}
-\frac{326673}{8\rho^3}
+\frac{580455}{16\rho}\bigg)\frac{1}{M^9}\nonumber\\
& &+O(M^{-10}) \nonumber\\
&=&\sum_{k=1}^{\infty}\frac{b_{k}}{M^k}.
\label{largem}
\end{eqnarray}

\section{Scaling in $1/M$ expansion and estimating the mass gap}
In this section, we first study the scaling behavior of  $\beta$ as a function of $M$.  Then, we try to evaluate the mass gap in the continuum limit via dilated $1/M$ series of $\beta$.  Finally, we discuss the same subject by choosing various combinations of variables relevant to the mass gap computation.

\subsection{Scaling and mass gap}
Assume the scaling form
\begin{equation}
\beta=A M^{-\alpha}(1+\cdots)
\label{scaling2}
\end{equation}
where $\cdots$ stands for the correction which vanishes in the $M\to 0$ limit.  The constant $A$ is dimensionless and directly connected to the value of the mass gap as we can see below:  From the definition of $\beta$, we obtain for small enough $M$
$$
A=2\lambda^{-1/2}a^{-3/2}M^{\alpha}.
$$
The right hand side is independent of $a$ and the quartic coupling converges to its continuum value $\lambda^{*}$ in the $a\to 0$ limit.   Now, let the dynamical mass $m$ in the continuum limit be defined conventionally by
\begin{equation}
m=\lim_{a\to 0}(\xi a)^{-1}.
\end{equation}
Then, since $M\to \xi^{-2}\sim m^2 a^2$ in the $a\to 0$ limit, 
$$
A=\lim_{a\to 0}2\lambda^{-1/2}a^{-3/2}M^{\alpha}=2(\lambda^{*})^{-1/2}m^{2\alpha}\lim_{a\to 0}a^{-3/2+2\alpha}.
$$
Thus, $a$ should disappear and then
\begin{equation}
\alpha=\frac{3}{4},
\label{alpha}
\end{equation}
and
\begin{equation}
m=\Big(\frac{A}{2}\Big)^{2/3}(\lambda^{*})^{1/3}.
\label{gap}
\end{equation}
The result (\ref{alpha}) is derived by assuming the generation of the finite dynamical mass.  We will show  by applying delta expansion that the assumption is actually confirmed in the large $M$ expansion of $\beta$.   Then, we turn to the evaluation of $A$ which cannot be guessed by such a dimensional argument.

\subsection{Dilation and delta expansion with respect to $M$}
To begin with let us see the behavior of $1/M$ series of $\beta$.  Figure 3 shows the plot of (\ref{largem}) at $2$nd and $9$th orders.  It is clear that the series breaks down around $M\sim 4$ or so and this implies the limitation of $1/M$ series.  The delta expansion drastically improves the utilities of $1/M$ series as we can see in the following.
\begin{figure}[h]
\begin{center}
\includegraphics[scale=0.6]{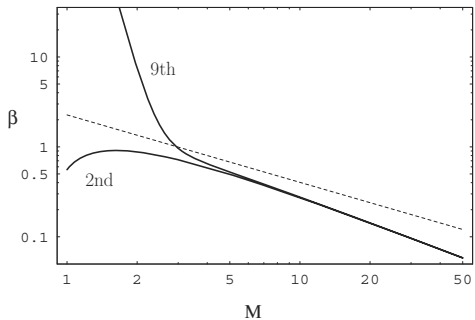}
\end{center}
\caption{Plots of $\beta$ as a function of $M$ at $2$nd and $9$th orders.  Dotted line represents the scaling behavior at small enough $M$.}
\end{figure} 

We dilate the scaling region of $M$ by introducing $\delta$ via $M\to M(1-\delta)$ in the function $\beta(M)$.  In the large $M$ expansion (\ref{largem}), this means simply the change $M^{-k}\to M^{-k}(1-\delta)^{-k}$ in respective terms.  As $\delta$ approaches to unity, $M^{-k}(1-\delta)^{-k}$ diverges to infinity, suggesting the violation of the large $M$ series.  This is the point where the expansion in $\delta$ comes into play.   We expand $M^{-k}(1-\delta)^{-k}$ such that $M^{-k}(1+k\delta+\frac{k(k+1)}{2!}\delta^2+\cdots)$ and truncate at a relevant order of $\delta$.  According to \cite{yam}, we adopt a prescription that $(M^{-1})^{i}\delta^{j}$ is of order $i+j$ and include it as a contribution to the full order $K$ as long as $i+j\leq K$.  Then we see that $(1-\delta)^{-k}$ should be expanded to $\delta^{K-k}$ and, by setting $\delta=1$ which means the infinite dilation, we find that $M^{-k}$ transforms as
\begin{equation}
M^{-k}\quad \to\quad \frac{K!}{k!(K-k)!}M^{-k}=\Big(
\begin{array}{c}
K \\
k
\end{array}
\Big)M^{-k}.
\end{equation}
Thus, we obtain the delta expansion of $\sum_{k=1}^{K}\frac{b_{k}}{M^k}=\beta_{K}$:
\begin{equation}
\beta_{K}\to D[\beta_{K}]= \sum_{k=1}^{K}\Big(
\begin{array}{c}
K \\
k
\end{array}
\Big)\frac{b_{k}}{M^k}.
\label{deltabeta}
\end{equation}
Figure 4 shows the plots of $D[\beta_{K}]$ at $K=2$ and $9$ and their asymptotic behaviors (shown by the dotted lines).   The asymptotic scaling behaviors plotted in Figure 4 are obtained as follows:  First note that the maximum order of $\delta$ in obtaining (\ref{deltabeta}) is $\delta^{K-1}$.  Accordingly the leading term in (\ref{scaling2}) should be expanded as $M^{-\alpha}(1-\delta)^{-\alpha}=M^{-\alpha}(1+\alpha\delta+\cdots)$ and truncated at $\delta^{K-1}$.  Then we find by setting $\delta=1$,
\begin{equation}
M^{-\alpha}\to M^{-\alpha}Z_{K}(\alpha),
\end{equation}
where
\begin{equation}
Z_{K}(\alpha)=\frac{\Gamma(K+\alpha)}{(K-1)!\Gamma(1+\alpha)}.
\end{equation}
Thus, the asymptotic behavior is also dilated and the expansion in $\delta$ gives
\begin{equation}
D[\beta_{K}]\sim A M^{-\alpha}\times Z_{K}(\alpha).
\end{equation}
It would be clear from Figure 4 that the power like behavior of the correct exponent (\ref{alpha}) is seen at some finite region of $M$.  For example at $K=9$, scaling behavior has emerged at the region of $M$ from $\sim 2$ to $\sim 5$.  Note that the region where the scaling is exhibited is not restricted to the neighborhood of $M=0$.  This is a characteristic feature of dilated functions.  The reason that the position of the dotted line shifts upwards as the order increases is that the factor $Z_{K}(3/4)$ is produced by the delta expansion.  The factor grows with $K$ as $Z_{K}(3/4)\sim K^{3/4}/\Gamma(1+3/4)$.
\begin{figure}[h]
\begin{center}
\includegraphics[scale=0.6]{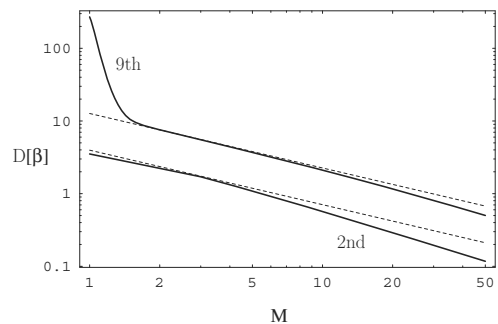}
\end{center}
\caption{Plots of $D[\beta_{K}]$ at $K=2$ and $9$.  Two dotted lines represent the scaling behavior at each orders.}
\end{figure} 

Having confirmed the scaling behavior, we like to show that the large $M$ series allows us to estimate the exponent $\alpha$.  For straightforward evaluation, it is convenient to deal with the function 
\begin{equation}
\frac{\partial \log \beta}{\partial \log M}=P(M).
\end{equation}
From (\ref{scaling2}), we find at small $M$,
\begin{equation}
P(M)=-\alpha+\cdots,
\label{p1}
\end{equation}
where $\cdots$ stands for the correction which tends to zero as $M\to 0$.  At large $M$ we obtain from (\ref{largem})
\begin{eqnarray}
P(M)&=&-1+ \Big(-\frac{1}{4\rho^2}+ 3\Big)\frac{1}{M}\nonumber\\
& &+\Big(-\frac{1}{6\rho^4}+\frac{5}{2\rho^2}-12\Big)\frac{1}{M^2}+\cdots\nonumber\\
&=&-1+\sum_{k=1}^{\infty}\frac{p_{k}}{M^k}.
\label{p2}
\end{eqnarray}
Now consider the dilation around $M=0$ with the amplification factor $(1-\delta)^{-1}$.  From (\ref{p1}), we find that $P(M(1-\delta))$ tends to the constant $-\alpha$ for all $M$ as $\delta$ goes to 1.   Hence, we examine whether the delta expansion of (\ref{p2}) exhibits the stationary behavior with the correct value of $-\alpha$.   Let us denote $P$ in $1/M$ expansion to the order $M^{-K}$ be $P_{K}$.  Then,  $D[P_{K}]$, the delta expanded series of $P_{K}$ at large $M$, reads
\begin{equation}
D[P_{K}]=-1+\sum_{k=1}^{K}\Big(
\begin{array}{c}
K \\
k
\end{array}
\Big)\frac{p_{k}}{M^k}.
\end{equation}
Figure 5 shows the plot of $D[P_{K}]$ for $K=2,5$ and $8$.  It is clearly seen that $D[P_{K}]$ indicates the correct value of $\alpha$.  Above orders $3$ or $4$, there appears a plateu and the width grows as the order of expansion increases.  This is a signal that the delta expansion is successfully working.
\begin{figure}[h]
\begin{center}
\includegraphics[scale=0.6]{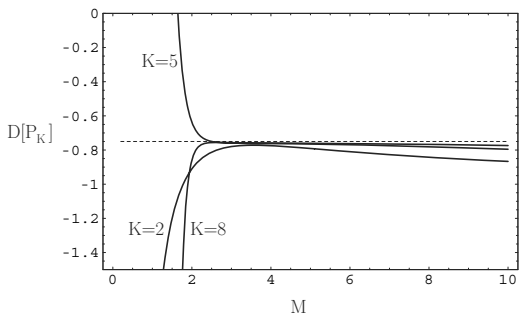}
\end{center}
\caption{Plots of $D[P_{K}]$ at $K=2,5$ and $8$.  Horizontal dotted line represents the value $-\alpha=-3/4$.}
\end{figure} 
The evaluation of $\alpha$ at even orders may be performed by noting that a typical value of $\alpha$ indicated at respective order is given by the stationary value of $-D[P_{K}]$.  The results at $K=2,4,6$ and $8$ are
\begin{equation}
0.77186,\quad 0.76239,\quad 0.75868,\quad 0.75548.
\end{equation}
These are all close to the exact value, $\alpha=3/4$.  Also at odd orders, evaluation is possible by selecting values at the minimum of $(\frac{\partial D[P_{K}]}{\partial M})^2$.

We can also perform the evaluation of the constant $A$ and the mass gap $m$.  For the purpose, it is convenient to consider $\log \beta$.  The function behaves at small $M$
\begin{equation}
\log \beta=\log A+\alpha\log M+\cdots,
\end{equation}
where $\cdots$ represents the correction and all terms in it vanish in the $M\to 0$ limit.  At large $M$, $\log \beta$ is written as
\begin{eqnarray}
\log\beta&=&-\log(\rho M)+\Big(-3+\frac{1}{4\rho^2}\Big)\frac{1}{M}\nonumber\\
& &+\Big(6+\frac{1}{12\rho^4}-\frac{5}{4\rho^2}\Big)\frac{1}{M^2}+\cdots.
\end{eqnarray}
By subtracting $\alpha\log M$ from $\log\beta$ we obtain the function $Q=\log \beta-\alpha\log M$ which converges to $\log A$ in the continuum limit;
\begin{eqnarray}
Q&=&-\log\rho-(1+\alpha)\log M+\Big(-3+\frac{1}{4\rho^2}\Big)\frac{1}{M}\nonumber\\
& &+\Big(6+\frac{1}{12\rho^4}-\frac{5}{4\rho^2}\Big)\frac{1}{M^2}+\cdots\nonumber\\
&=&-\log\rho-(1+\alpha)\log M+\sum_{k=1}^{\infty}\frac{q_{k}}{M^{k}}.
\end{eqnarray}
$\log M$, the leading term in $Q$, may be considered as the 0th order in $1/M$ expansion.  Hence, at the full order $K$, it is natural to expand $\log M(1-\delta)$ to $\delta^{K}$.  Then, we obtain by setting $\delta=1$, 
$$
\log M\to \log M-\sum_{k=1}^{K}\frac{1}{k}.
$$
Hence, denoting $Q$ to the order $K$ as $Q_{K}$,
\begin{equation}
D[Q_{K}]=-\log\rho-(1+\alpha)\Big(\log M-\sum_{k=1}^{K}\frac{1}{k}\Big)+\sum_{k=1}^{K}\Big(
\begin{array}{c}
K \\
k
\end{array}
\Big)\frac{q_{k}}{M^{k}}.
\end{equation}
We set $\alpha=3/4$ to evaluate the amplitude $A$.  
Figure 6 shows the plots of $D[Q_{K}]$ at $K=2, 5$ and $8$.  At every orders $D[Q_{K}]$ has a plateu at which the values of the function are around $\sim 0.8$.  For example, taking an extremum value as a  typical value, we obtain the following values of $\log A$ at orders $1$st, $3$rd, $5$th and $7$th, respectively:
$$
0.7903,\quad 0.8080,\quad 0.8137,\quad 0.8162.
$$
The width of the plateu grows as the order increases and thus the delta expansion method is surely successful.
\begin{figure}[h]
\begin{center}
\includegraphics[scale=0.6]{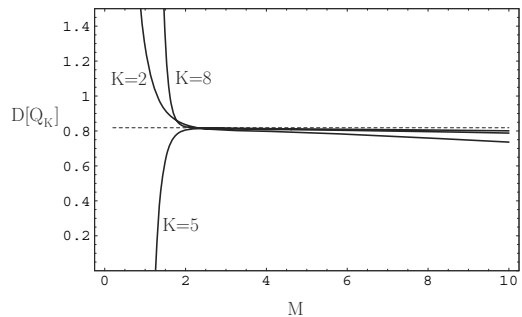}
\end{center}
\caption{Plots of $D[Q_{K}]$ at $K=2, 5$ and $8$.  Horizontal dotted line represents $\log A=0.81842\cdots$ which is taken from \cite{nev}.}
\end{figure} 
Taking typical values of $\log A$ as listed above, we obtain from (\ref{gap}) the following approximation of the mass gap in the continuum limit,
\begin{equation}
\frac{m}{(\lambda^{*})^{1/3}}=1.06696,\quad 1.07958,\quad 1.08376,\quad 1.08554.
\end{equation}
We can say that the approximation is in good agreement with the rigorous value,  $m/(\lambda^{*})^{1/3}=1.087096\cdots$ \cite{nev}.

\subsection{The comparison of various choices of basic parameter under the delta expansion}
Up to now, we have argued the scaling behavior of $\beta$ as a function of $M$.   There are other choices of two quantities such that the mutual dependence at the scaling region gives the mass gap in the continuum limit.  Furthermore, the choice of the basic variable to which the dilation is applied gives additional variations in our approach.  In this subsection, we report the results in other various cases.

One natural basic parameter is $\beta$, since it appears in the action and used as an expansion parameter in the hopping expansion.  Thus we study the scaling of $\xi^{-1}$ and $M^{-1}$ as functions of $\beta$.    Since $\beta\to \infty$ in the continuum limit, we make dilation around $\beta=\infty$ by shifting $\beta\to \beta/(1-\delta)$.  

First consider $\xi^{-1}(\beta)$.   We examine whether it scales as $\xi^{-1}\sim A^{\frac{1}{2\alpha}}\beta^{-\frac{1}{2\alpha}}=A^{\frac{2}{3}}\beta^{-\frac{2}{3}}$ by performing the delta expansion.   At small $\beta$, the delta expansion of $\xi^{-1}$ to order $K$ reads from (\ref{xibetascale})
\begin{equation}
D[\xi^{-1}]=-\log(\rho\beta)-\sum_{k=1}^{K}\frac{1}{k}+\sum_{k=1}^{K}\Big(
\begin{array}{c}
K \\
k
\end{array}\Big)
z_{k}\beta^k.
\label{deltaxi}
\end{equation}
At large $\beta$, we note that $D[\xi^{-1}]\sim A^{\frac{2}{3}}\beta^{-\frac{2}{3}}Z_{K}(\frac{2}{3})$ and the scaling behavior of $D[\xi^{-1}]/Z_{K}(2/3)$ agrees with that of $\xi^{-1}$.
In  Figure 7, we have plotted the functions $D[\xi^{-1}]/Z_{K}(2/3)$ at small $\beta$ and the asymptotic scaling behavior, $\xi^{-1}\sim A^{2/3}(\beta^{-1})^{2/3}$.   Though the functions show rough scaling at the region of  $1/\beta\sim O(1)$, they oscillate around the rigorous scaling.  In this case, we find small $\beta$ expansion of  $D[\xi^{-1}]/Z_{K}(2/3)$ is not sufficient  for further quantitative use.
\begin{figure}[h]
\begin{center}
\includegraphics[scale=0.60]{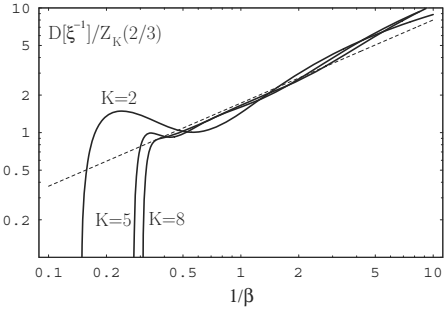}
\end{center}
\caption{Plots of $D[\xi^{-1}]/Z_{K}(2/3)$ at $K=2, 5$ and $8$.  Dotted line represents the scaling behavior of $\xi^{-1}$.}
\end{figure} 
Next we study the scaling of $M$ as a function of $\beta$.  From (\ref{smallb}), we have
 $D[M^{-1}]=\sum_{k=1}^{K}\Big(
\begin{array}{c}
K \\
k
\end{array}
\Big)m_{k}\beta^k$ at small $\beta$.  At large $\beta$, $D[M^{-1}]\sim A^{-1/\alpha}\times \beta^{1/\alpha}Z_{K}(4/3)=A^{-4/3}\beta^{4/3}\times Z_{K}(4/3)$.   Plots of $D[M^{-1}]/Z_{K}(4/3)$ at $K=3, 6$ and $9$ and the asymptotic scaling of $M^{-1}$ are shown in Figure 8.   From Figure 8 it seems that the scaling behavior is exhibited in small $\beta$ series.
\begin{figure}
\includegraphics[scale=0.62]{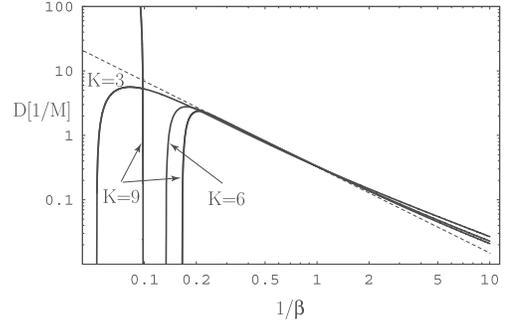}
\caption{$D[M^{-1}]/Z_{K}(4/3)$ at orders of $3$rd, $6$th and $9$th.   At $9$th order, the amplitude of oscillation becomes very large at $\beta \gg 1$.}
\end{figure}
Then we turn to the evaluation of the exponent $1/\alpha=4/3$ as in the same manner of the previous subsection.   Let us define the function $P_{\beta}=\frac{\partial \log 1/M}{\partial\log \beta}$ and apply the delta expansion to $P_{\beta}$.   Figure 9 shows the plot of $D[P_{\beta,K}]$.  
The result indicates the correct value of $1/\alpha=4/3$.  However, due to the oscillatory nature, the explicit evaluation of the exponent is not a straightforward task.  
We conclude that considering $\xi^{-1}(\beta)$ and $M^{-1}(\beta)$ gives us rough and good scaling respectively but they are not  the best choice for our approach. 
\begin{figure}[hb]
\begin{center}
\includegraphics[scale=0.5]{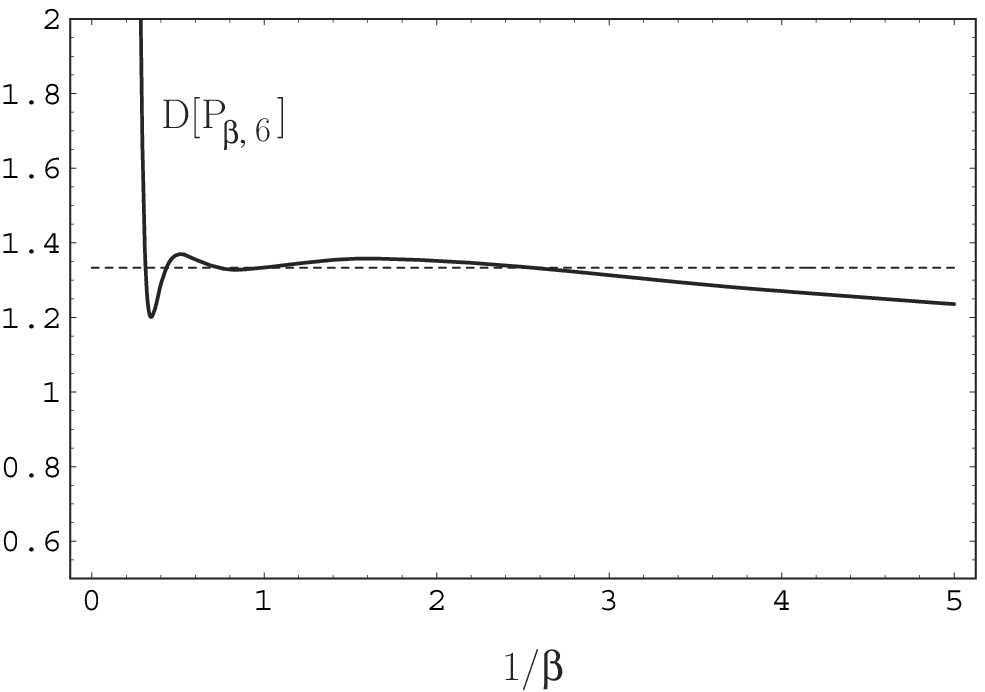}
\includegraphics[scale=0.5]{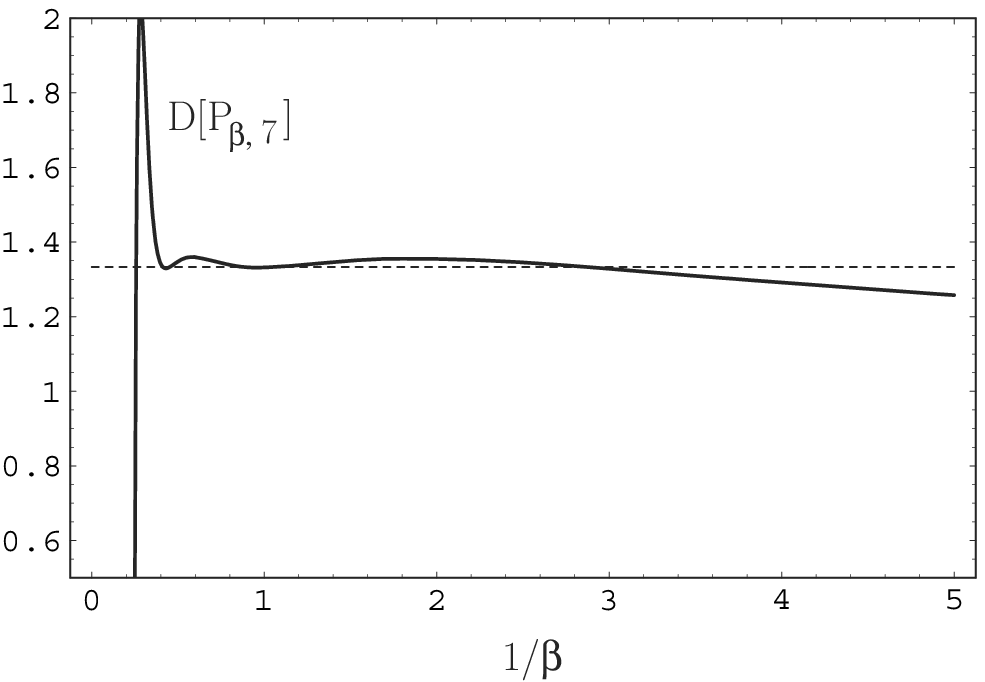}
\includegraphics[scale=0.5]{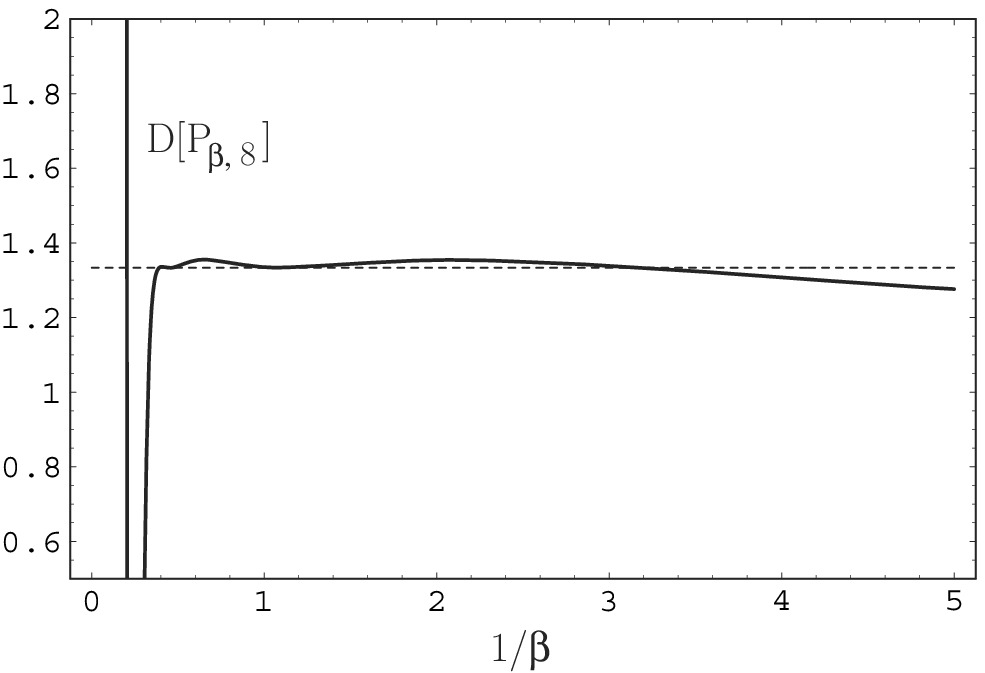}
\end{center}
\caption{Plots of $D[P_{\beta, K}]$ at $K=6,7,8$.  Horizontal dotted line represents $P_{\beta}|_{\beta=\infty}=1/\alpha=4/3$.  Dotted line in each graph represents the value $\frac{1}{\alpha}=\frac{4}{3}$. }
\end{figure} 

As the last case, let us consider $\beta(\xi)$.  To perform our method, one needs to invert (\ref{corr}).  This may be done by expressing $M^{-1}$ in terms of $\xi$ and substituting the result into (2.10).  Then one finds that $\beta$ is expanded as a series of $e^{-1/\xi}$ such that $\beta=\frac{1}{\rho}e^{-1/\xi}+(\frac{1}{4\rho^3}-\frac{1}{\rho})e^{-2/\xi}+(\frac{11}{96\rho^5}-\frac{1}{\rho^3}+\frac{3}{2\rho})e^{-3/\xi}+\cdots$.  
  The expansion parameter becomes $e^{-1/\xi}$ and the continuum limit is taken out in it giving $e^{-1/\xi}\to 1$.  The divergence of the continuum limit appears only in the whole series.  In other words, expansion of $\beta$ at small $\xi$ is like a "low temperature expansion".  Since the new delta expansion has been applied only to the "high temperature expansion" in basic parameter, we are in a situation different from the previous cases.   Then let us proceed in a formal way.  We dilate the region of $\xi$ around $\xi=\infty$ by shifting $\xi\to \xi/(1-\delta)$ and then carry out expansion in $\delta$.  The result is, however, found to be poor and $\beta(\xi)$ is not enough for the present purpose.  We would like to mention that this problem was also reported at the large ${\cal N}$ anharmonic oscillator \cite{cit}.  Thus we conclude that the choice of $\xi$ as a basic parameter and considering $\beta(\xi)$ is not adequate in our approach.

\section{Conclusion}
The mass gap in the continuum limit is investigated by seeking the relationship between $M$ and $\beta$ or $\xi$ and $\beta$ at scaling.  In our study based upon the delta expansion, the former pair is much convenient than the later.   
In fact, the superiority of $M$ over $\xi^{-1}$ is apparent in the case of the harmonic oscillator.   The small $\beta$ expansion of $\xi^{-1}$ and $M$ reads respectively
$$
\xi^{-1}=-\log\frac{\beta}{2}+\beta-\frac{3}{4}\beta^2+\frac{5}{6}\beta^3-\frac{19}{32}\beta^4+\cdots
$$
and
$$
M^{-1}=\frac{\beta}{2}.
$$
Here note that $\beta=\frac{2}{(am)^2}$ where $m$ denotes the current mass.  
For $\xi^{-1}$ the expansion becomes an infinite series and the inversion, too.  On the contrary, for $M^{-1}$, the expansion terminates just at the first order of $\beta$ and the result is exact.  Thus, our work suggests that the mass variable in the momentum space is more suitable than the correlation length defined on the lattice space.

To conclude the present work, we have applied the dilation and expansion in dilation parameter to the anharmonic oscillator at strong coupling.  The mass gap calculation was studied by choosing variables, the mass in the momentum space, the correlation length and the hopping parameter, as the parameter to which the dilation is applied.  We found that the mass in the momentum space produced the best performance and the scaling of $\beta(M)$ at small enough $M$ was clearly seen in the $\delta$-expanded $1/M$ expansion at rather wide region of shifted $M$.  The computation of the mass gap in the continuum limit was also successfully done.

\appendix
\section{Expansion of $\gamma_{j}$}
The expansion of $\gamma_{j}$ in powers of $\beta$ can be carried out as follows:  First consider the expansion of $h_{0}$ given by $h_{0}=\int_{-\infty}^{\infty}d\varphi e^{-(\beta\varphi^2+\varphi^{4})}$.  We expand $e^{-\beta\varphi^2}$ as $\sum_{l=0}^{\infty}\frac{(-\beta)^l}{l!}\varphi^{2l}$ and evaluate the integral,
$$
\int_{-\infty}^{\infty}d\varphi e^{-\varphi^4}\times\varphi^{2l}=\frac{1}{2}\Gamma\Big(\frac{l}{2}+\frac{1}{4}\Big).
$$
Thus, we have
$$
h_{0}=\sum_{l=0}^{\infty}\frac{(-\beta)^l}{2 l!}\Gamma\Big(\frac{l}{2}+\frac{1}{4}\Big).
$$
Then, by using $\gamma_{j}=\frac{1}{h_{0}}\Big(-\frac{\partial }{\partial\beta}\Big)^{j}h_{0}$, we obtain the expansion of $\gamma_{j}$ in powers of $\beta$.  The results become simple if one reduces the argument of $\Gamma$ functions by the formula, $\Gamma(z+1)=z\Gamma(z)$.  Actually, the coefficients of $\beta^{l}$ ($l=0,1,2,\cdots)$ are found to be written in $\Gamma(3/4)/\Gamma(1/4)=\rho$.

To $8$th order we need $\gamma_{j}$ for $j=1$ to $5$.  The results of the expansion are summarized below.  
\begin{eqnarray}
\gamma_{1}&=&\rho+\Big(-\frac{1}{4}+\rho^2\Big)\beta+\rho^3\beta^2+\Big(-\frac{1}{48}+\rho^4\Big)\beta^3\nonumber\\
& &+\Big(-\frac{\rho}{96}+\rho^5\Big)\beta^4+\Big(-\frac{\rho^2}{80}+\rho^6\Big)\beta^5\nonumber\\
& &+\Big(-\frac{7\rho^3}{480}+\rho^7\Big)\beta^6+\Big(\frac{1}{16128}-\frac{\rho^4}{60}+\rho^8\Big)\beta^7\nonumber\\
& & + 
\Big(\frac{\rho}{14336}-\frac{3\rho^5}{160}+\rho^9\Big)\beta^8+O(\beta^9)\nonumber\\
\gamma_{2}&=&\frac{1}{4}-\frac{\rho}{2}\beta+\Big(\frac{1}{8}-\frac{\rho^2}{2}\Big)\beta^2-\frac{\rho^3}{2}\beta^3+\Big(\frac{1}{96}-\frac{\rho^4}{2}\Big)\beta^4\nonumber\\
& &+\Big(\frac{\rho}{192}-\frac{\rho^5}{2}\Big)\beta^5+\Big(\frac{\rho^2}{160}-\frac{\rho^6}{2}\Big)\beta^6\nonumber\\
& &\Big(\frac{7\rho^3}{960}-\frac{\rho^7}{2}\Big)\beta^7 + 
\Big(-\frac{1}{32256}+\frac{\rho^4}{120}-\frac{\rho^8}{2}\Big)\beta^8\nonumber\\
& &+O(\beta^9) \nonumber\\
\gamma_{3}&=&\frac{3\rho}{4}+\Big(-\frac{5}{16}+\frac{3\rho^2}{4}\Big)\beta+\Big(\frac{\rho}{4}+\frac{3\rho^3}{4}\Big)\beta^2\nonumber\\
& &+\Big(-\frac{5}{64}+\frac{\rho^2}{4}+\frac{3\rho^4}{4}\Big)\beta^3+\Big(-\frac{\rho}{128}+\frac{\rho^3}{4}+\frac{3\rho^5}{4}\Big)\beta^4\nonumber\\
& &+\Big(-\frac{1}{192}-\frac{3\rho^2}{320}+\frac{\rho^4}{4}+\frac{3\rho^6}{4}\Big)\beta^5\nonumber\\
& &+\Big(-\frac{\rho}{384}-\frac{7\rho^3}{640}+\frac{\rho^5}{4}+\frac{3\rho^7}{4}\Big)\beta^6\nonumber\\
& &+\left(\frac{1}{21504}-\frac{\rho^2}{320}-\frac{\rho^4}{80}+\frac{\rho^6}{4}+\frac{3\rho^8}{4}\right)\beta^7\nonumber\\
& & + 
\left(\frac{3\rho}{57344}-\frac{7\rho^3}{1920}-\frac{9\rho^5}{640}+\frac{\rho^7}{4}+\frac{3\rho^9}{4}\right)\beta^8+O(\beta^9) \nonumber\\
\gamma_{4}&=&\frac{5}{16}-\rho\beta+\Big(\frac{5}{16}-\rho^2\Big)\beta^2+\Big(-\frac{\rho}{8}-\rho^3\Big)\beta^3\nonumber\\
& &+\Big(\frac{5}{96}-\frac{\rho^2}{8}-\rho^4\Big)\beta^4+\Big(\frac{\rho}{96}-\frac{\rho^3}{8}-\rho^5\Big)\beta^5\nonumber\\
& &+\Big(\frac{1}{384}+\frac{\rho^2}{80}-\frac{\rho^4}{8}-\rho^6\Big)\beta^6\nonumber\\
& &+\left(\frac{\rho}{768}+\frac{7\rho^3}{480}-\frac{\rho^5}{8}-\rho^7\right)\beta^7\nonumber\\
& & + 
\left(-\frac{1}{16128}+\frac{\rho^2}{640}+\frac{\rho^4}{60}-\frac{\rho^6}{8}-\rho^8\right)\beta^8+O(\beta^9)\nonumber\\
\gamma_{5}&=&
\frac{21\rho}{16} + 
\left(-\frac{45}{64}+\frac{21\rho^2}{16}\right)\beta + 
\left(\frac{15\rho}{16}+\frac{21\rho^3}{16}\right)\beta^2\nonumber\\
& & + 
\left(-\frac{75}{256}+\frac{15\rho^2}{16}+\frac{21\rho^4}{16}\right)\beta^3\nonumber\\
& & + 
\left(\frac{25\rho}{512}+\frac{15\rho^3}{16}+\frac{21\rho^5}{16}\right)\beta^4\nonumber\\
& & + 
\left(-\frac{9}{256}+\frac{59\rho^2}{1280}+\frac{15\rho^4}{16}+\frac{21\rho^6}{16}\right)\beta^5\nonumber\\
& & + 
\left(-\frac{5\rho}{512}+\frac{111\rho^3}{2560}+\frac{15\rho^5}{16}+\frac{21\rho^7}{16}\right)\beta^6\nonumber\\
& & + 
\left(-\frac{5}{4096}-\frac{3\rho^2}{256}+\frac{13\rho^4}{320}+\frac{15\rho^6}{16}+\frac{21\rho^8}{16}\right)\beta^7\nonumber\\
& & + 
\left(-\frac{55\rho}{98304}-\frac{7\rho^3}{512}+\frac{97\rho^5}{2560}+\frac{15\rho^7}{16}+\frac{21\rho^9}{16}\right)\beta^8\nonumber\\
& &+O(\beta^9)
\label{gamma}
\end{eqnarray}

\section{Results of $c_{k}$}
We have computed the expansion of the two point correlation function $\la\varphi_{0}\varphi_{n}\ra$ to the $8$th order of the hopping term with the help of the graphical representation.  The total lattice size $L$ is present in the numerator but all the contributions dependent on $L^{m}$ $(m=1,2,3,\cdots)$ are cancelled by the contributions of the partition function.  Thus, only the contributions of the order $L^{0}$ is left and they are collected in terms of $n$.  The coefficients of $n^{l}$ ($l=0,1,2,3,4$) are written below as a series of the hopping parameter $\beta$.
\begin{eqnarray}
c_{0}&=&\gamma_{1}+\beta^2(-\gamma_{1}^3+\gamma_{1}\gamma_{2})+\beta^4\Big(\frac{3\gamma_{1}^5}{2}-\frac{9\gamma_{1}^3\gamma_{2}}{4}+\frac{2\gamma_{1}\gamma_{2}^2}{3}\nonumber\\
& &-\frac{\gamma_{2}^3}{6\gamma_{1}}+\frac{\gamma_{2}^4}{36\gamma_{1}^3}+\frac{\gamma_{2}\gamma_{3}}{4}-\frac{\gamma_{2}^2\gamma_{3}}{36\gamma_{1}^2}\Big)+\beta^6\Big(-\frac{5\gamma_{1}^7}{2}\nonumber\\
& &+5\gamma_{1}^5\gamma_{2}-\frac{11\gamma_{1}^3\gamma_{2}^2}{4}+\frac{5\gamma_{1}\gamma_{2}^3}{8}-\frac{7\gamma_{2}^4}{36\gamma_{1}}+\frac{\gamma_{2}^5}{12\gamma_{1}^3}\nonumber\\
& &-\frac{\gamma_{2}^6}{54\gamma_{1}^5}-\frac{\gamma_{1}^2\gamma_{2}\gamma_{3}}{2}+\frac{5\gamma_{2}^2\gamma_{3}}{18}-\frac{\gamma_{2}^3\gamma_{3}}{8\gamma_{1}^2}+\frac{\gamma_{2}^4\gamma_{3}}{36\gamma_{1}^4}\nonumber\\
& &+\frac{7\gamma_{1}\gamma_{3}^2}{180}+\frac{7\gamma_{2}\gamma_{3}^2}{360\gamma_{1}}-\frac{7\gamma_{2}^2\gamma_{3}^2}{1080\gamma_{1}^3}+\frac{\gamma_{2}^2\gamma_{4}}{72\gamma_{1}}+\frac{\gamma_{3}\gamma_{4}}{90}\nonumber\\
& &-\frac{\gamma_{2}\gamma_{3}\gamma_{4}}{360\gamma_{1}^2}\Big)+\beta^8
\bigg(
\frac{35\gamma_1^9}{8}-\frac{175\gamma_1^7\gamma_2}{16}+\frac{35\gamma_1^5\gamma_2^2}{4}\nonumber\\
& &-\frac{35\gamma_1^3\gamma_2^3}{12}+\frac{71\gamma_1\gamma_2^4}{96}-
\frac{\gamma_2^5}{4\gamma_1}+\frac{43\gamma_2^6}{432\gamma_1^3}-\frac{5\gamma_2^7}{108\gamma_1^5}\nonumber\\
& &+\frac{5\gamma_2^8}{432\gamma_1^7}+\frac{25\gamma_1^4\gamma_2\gamma_3}{24}-
\frac{25\gamma_1^2\gamma_2^2\gamma_3}{24}+\frac{35\gamma_2^3\gamma_3}{96}\nonumber\\
& &-\frac{71\gamma_2^4\gamma_3}{432\gamma_1^2}+\frac{13\gamma_2^5\gamma_3}{144\gamma_1^4}-\frac{5\gamma_2^6\gamma_3}{216\gamma_1^6}-
\frac{7\gamma_1^3\gamma_3^2}{60}+\frac{7\gamma_1\gamma_2\gamma_3^2}{240}\nonumber\\
& &+\frac{359\gamma_2^2\gamma_3^2}{8640\gamma_1}-\frac{41\gamma_2^3\gamma_3^2}{1080\gamma_1^3}+
\frac{\gamma_2^4\gamma_3^2}{90\gamma_1^5}+\frac{\gamma_2\gamma_3^3}{864\gamma_1^2}-\frac{\gamma_2^2\gamma_3^3}{1080\gamma_1^4}\nonumber\\
& &+\frac{\gamma_3^4}{14400\gamma_1^3}-\frac{5\gamma_1\gamma_2^2\gamma_4}{192}+
\frac{\gamma_2^3\gamma_4}{144\gamma_1}-\frac{\gamma_2^4\gamma_4}{216\gamma_1^3}\nonumber\\
& &-\frac{\gamma_1^2\gamma_3\gamma_4}{48}+\frac{7\gamma_2\gamma_3\gamma_4}{288}-\frac{7\gamma_2^2\gamma_3\gamma_4}{1080\gamma_1^2}+
\frac{\gamma_2^3\gamma_3\gamma_4}{360\gamma_1^4}\nonumber\\
& &+\frac{\gamma_3^2\gamma_4}{720\gamma_1}-\frac{\gamma_2\gamma_3^2\gamma_4}{1080\gamma_1^3}+\frac{3\gamma_1\gamma_4^2}{2240}+\frac{\gamma_2\gamma_4^2}{840\gamma_1}-
\frac{\gamma_2^2\gamma_4^2}{2520\gamma_1^3}\nonumber\\
& &+\frac{\gamma_2\gamma_3\gamma_5}{864\gamma_1}-\frac{\gamma_3^2\gamma_5}{14400\gamma_1^2}+\frac{\gamma_4\gamma_5}{4032}-\frac{\gamma_2\gamma_4\gamma_5}{15120\gamma_1^2}
\bigg)+O(\beta^{10})\nonumber\\
c_{1}&=&\beta^2\Big(-\frac{\gamma_{1}^3}{2}+\frac{\gamma_{2}^2}{6\gamma_{1}}\Big)+\beta^4\Big(\frac{7\gamma_{1}^5}{8}-\frac{3\gamma_{1}^3\gamma_{2}}{4}-\frac{5\gamma_{1}\gamma_{2}^2}{24}\nonumber\\
& &+\frac{\gamma_{2}^3}{6\gamma_{1}}-\frac{\gamma_{2}^4}{24\gamma_{1}^3}+\frac{\gamma_{2}^2\gamma_{3}}{36\gamma_{1}^2}+\frac{\gamma_{3}^2}{120\gamma_{1}}\Big)+\beta^6\Big(-\frac{37\gamma_{1}^7}{24}\nonumber\\
& &+\frac{9\gamma_{1}^5\gamma_{2}}{4}-\frac{17\gamma_{1}^3\gamma_{2}^2}{48}-\frac{\gamma_{1}\gamma_{2}^3}{3}+\frac{5\gamma_{2}^4}{36\gamma_{1}}-\frac{5\gamma_{2}^5}{72\gamma_{1}^3}\nonumber\\
& &+\frac{13\gamma_{2}^6}{648\gamma_{1}^5}-\frac{\gamma_{1}^2\gamma_{2}\gamma_{3}}{6}-\frac{\gamma_{2}^2\gamma_{3}}{72}+\frac{5\gamma_{2}^3\gamma_{3}}{72\gamma_{1}^2}-\frac{5\gamma_{2}^4\gamma_{3}}{216\gamma_{1}^4}\nonumber\\[20pt]
& &-\frac{7\gamma_{1}\gamma_{3}^2}{720}+\frac{\gamma_{2}\gamma_{3}^2}{120\gamma_{1}}+\frac{\gamma_{2}^2\gamma_{3}^2}{2160\gamma_{1}^3}+\frac{\gamma_{2}\gamma_{3}\gamma_{4}}{360\gamma_{1}^2}+\frac{\gamma_{4}^2}{5040\gamma_{1}}\Big)\nonumber\\
& &+\beta^8
\bigg(
\frac{533\gamma_1^9}{192}-\frac{535\gamma_1^7\gamma_2}{96}+\frac{253\gamma_1^5\gamma_2^2}{96}+\frac{29\gamma_1^3\gamma_2^3}{96}\nonumber\\
& &-\frac{457\gamma_1\gamma_2^4}{1152}+
\frac{5\gamma_2^5}{32\gamma_1}-\frac{175\gamma_2^6}{2592\gamma_1^3}+\frac{47\gamma_2^7}{1296\gamma_1^5}-\frac{19\gamma_2^8}{1728\gamma_1^7}\nonumber\\
& &+\frac{15\gamma_1^4\gamma_2\gamma_3}{32}-
\frac{67\gamma_1^2\gamma_2^2\gamma_3}{288}-\frac{25\gamma_2^3\gamma_3}{288}+\frac{61\gamma_2^4\gamma_3}{864\gamma_1^2}\nonumber\\
& &-\frac{47\gamma_2^5\gamma_3}{864\gamma_1^4}+\frac{47\gamma_2^6\gamma_3}{2592\gamma_1^6}-
\frac{\gamma_1^3\gamma_3^2}{72}-\frac{43\gamma_1\gamma_2\gamma_3^2}{1440}+\frac{2\gamma_2^2\gamma_3^2}{135\gamma_1}\nonumber\\
& &+\frac{\gamma_2^3\gamma_3^2}{108\gamma_1^3}-\frac{11\gamma_2^4\gamma_3^2}{2160\gamma_1^5}+
\frac{\gamma_2\gamma_3^3}{480\gamma_1^2}-\frac{\gamma_2^2\gamma_3^3}{2592\gamma_1^4}-\frac{\gamma_3^4}{9600\gamma_1^3}\nonumber\\
& &-\frac{5\gamma_1\gamma_2^2\gamma_4}{576}+\frac{\gamma_2^4\gamma_4}{432\gamma_1^3}-
\frac{\gamma_1^2\gamma_3\gamma_4}{144}-\frac{\gamma_2\gamma_3\gamma_4}{720}\nonumber\\
& &+\frac{\gamma_2^2\gamma_3\gamma_4}{216\gamma_1^2}-\frac{\gamma_2^3\gamma_3\gamma_4}{432\gamma_1^4}+\frac{\gamma_2\gamma_3^2\gamma_4}{2160\gamma_1^3}-
\frac{\gamma_1\gamma_4^2}{4480}+\frac{\gamma_2\gamma_4^2}{5040\gamma_1}\nonumber\\
& &+\frac{\gamma_2^2\gamma_4^2}{7560\gamma_1^3}+\frac{\gamma_3^2\gamma_5}{14400\gamma_1^2}+\frac{\gamma_2\gamma_4\gamma_5}{15120\gamma_1^2}+\frac{\gamma_5^2}{362880\gamma_1}
\bigg)\nonumber\\
& &+O(\beta^{10})\nonumber\\
c_{2}&=&\beta^4\Big(\frac{\gamma_{1}^5}{8}-\frac{\gamma_{1}\gamma_{2}^2}{12}+\frac{\gamma_{2}^4}{72\gamma_{1}^3}\Big)+\beta^6\Big(-\frac{5\gamma_{1}^7}{16}+\frac{\gamma_{1}^5\gamma_{2}}{4}\nonumber\\
& &+\frac{\gamma_{1}^3\gamma_{2}^2}{6}-\frac{\gamma_{1}\gamma_{2}^3}{8}+\frac{\gamma_{2}^5}{72\gamma_{1}^3}-\frac{\gamma_{2}^6}{144\gamma_{1}^5}-\frac{\gamma_{2}^2\gamma_{3}}{72}\nonumber\\
& &+\frac{\gamma_{2}^4\gamma_{3}}{216\gamma_{1}^4}-\frac{\gamma_{1}\gamma_{3}^2}{240}+\frac{\gamma_{2}^2\gamma_{3}^2}{720\gamma_{1}^3}\Big)+\beta^8
\bigg(
\frac{251\gamma_1^9}{384}\nonumber\\
& &-\frac{15\gamma_1^7\gamma_2}{16}-\frac{13\gamma_1^5\gamma_2^2}{576}+\frac{37\gamma_1^3\gamma_2^3}{96}-\frac{43\gamma_1\gamma_2^4}{384}+
\frac{\gamma_2^5}{144\gamma_1}\nonumber\\
& &+\frac{41\gamma_2^6}{5184\gamma_1^3}-\frac{\gamma_2^7}{108\gamma_1^5}+\frac{119\gamma_2^8}{31104\gamma_1^7}+\frac{5\gamma_1^4\gamma_2\gamma_3}{96}\nonumber\\
& &+
\frac{\gamma_1^2\gamma_2^2\gamma_3}{48}-\frac{7\gamma_2^3\gamma_3}{144}+\frac{5\gamma_2^4\gamma_3}{864\gamma_1^2}+\frac{7\gamma_2^5\gamma_3}{864\gamma_1^4}-\frac{\gamma_2^6\gamma_3}{216\gamma_1^6}\nonumber\\
& &+
\frac{23\gamma_1^3\gamma_3^2}{2880}-\frac{\gamma_1\gamma_2\gamma_3^2}{160}-\frac{19\gamma_2^2\gamma_3^2}{8640\gamma_1}+\frac{\gamma_2^3\gamma_3^2}{720\gamma_1^3}+\frac{\gamma_2^4\gamma_3^2}{8640\gamma_1^5}\nonumber\\
& &+
\frac{\gamma_2^2\gamma_3^3}{4320\gamma_1^4}+\frac{\gamma_3^4}{28800\gamma_1^3}-\frac{\gamma_2\gamma_3\gamma_4}{720}+\frac{\gamma_2^3\gamma_3\gamma_4}{2160\gamma_1^4}-\frac{\gamma_1\gamma_4^2}{10080}\nonumber\\
& &+\frac{\gamma_2^2\gamma_4^2}{30240\gamma_1^3}
\bigg)+O(\beta^{10})\nonumber\\
c_{3}&=&\beta^6\Big(-\frac{\gamma_{1}^7}{48}+\frac{\gamma_{1}^3\gamma_{2}^2}{48}-\frac{\gamma_{2}^4}{144\gamma_{1}}+\frac{\gamma_{2}^6}{1296\gamma_{1}^5}\Big)\nonumber\\
& &
+\beta^8
\bigg(
\frac{13\gamma_1^9}{192}-\frac{5\gamma_1^7\gamma_2}{96}-\frac{11\gamma_1^5\gamma_2^2}{192}+\frac{\gamma_1^3\gamma_2^3}{24}+\frac{\gamma_1\gamma_2^4}{96}\nonumber\\
& &-
\frac{\gamma_2^5}{96\gamma_1}+\frac{11\gamma_2^6}{5184\gamma_1^3}+\frac{\gamma_2^7}{1296\gamma_1^5}-\frac{\gamma_2^8}{1728\gamma_1^7}+\frac{\gamma_1^2\gamma_2^2\gamma_3}{288}\nonumber\\
& &-
\frac{\gamma_2^4\gamma_3}{432\gamma_1^2}+\frac{\gamma_2^6\gamma_3}{2592\gamma_1^6}+\frac{\gamma_1^3\gamma_3^2}{960}-\frac{\gamma_2^2\gamma_3^2}{1440\gamma_1}+\frac{\gamma_2^4\gamma_3^2}{8640\gamma_1^5}
\bigg)\nonumber\\
& &+O(\beta^{10})\nonumber\\
c_{4}&=&\beta^8\bigg(\frac{\gamma_{1}^9}{384}-\frac{\gamma_{1}^5\gamma_{2}^2}{288}+\frac{\gamma_{1}\gamma_{2}^4}{576}-\frac{\gamma_{2}^6}{2592\gamma_{1}^3}+\frac{\gamma_{2}^8}{31104\gamma_{1}^7}\bigg)\nonumber\\
& &+O(\beta^{10})
\end{eqnarray}


\begin{thebibliography}{99}
\bibitem{kappa} The expansion parameter $\beta$ in our method is twice of $\kappa$ which is used in the conventional hopping expansion.
\end{thebibliography}
\end{document}